\documentclass[reprint,prc,amsmath,amssymb,aps]{revtex4-1}
\usepackage{natbib}
\setcitestyle{square, comma, numbers,sort&compress}
\usepackage{graphicx}% Include figure files
\usepackage{dcolumn}% Align table columns on decimal point
\usepackage{bm}% bold math
\usepackage{hyperref}% add hypertext capabilities
\usepackage[mathlines]{lineno}% Enable numbering of text and display math
\usepackage[toc,page]{appendix}
\usepackage{todonotes}
\usepackage{theorem}
\usepackage{color}
\usepackage{rotating}
\usepackage{epstopdf}
\usepackage{epsfig}
\usepackage{notoccite}
\usepackage{multirow}
\usepackage{tikz}
\usepackage{array}
\usepackage{ltablex}
\usepackage{upgreek}
\usepackage{titlesec}
\usepackage{tikz}
\usepackage{amsmath}
\usetikzlibrary{shapes,arrows}

\begin{document}

\title{Three-dimensional femtoscopy for proton pairs in heavy-ion collisions}

\author{Adam Kisiel}
\email{adam.kisiel@pw.edu.pl}
\affiliation{Faculty of Physics, Warsaw University of Technology\\
 ul. Koszykowa 75, 00-662 Warsaw, Poland}


%\input{abstract.tex}
\begin{abstract}
Femtoscopy, a technique of measuring the size and the dynamics of the system created in heavy-ion collisions is used extensively in experiments at RHIC, LHC, SPS, and the FAIR/GSI. Analysis for pairs of pions is most common, due to their large abundance in the collisions, as well as well-understood theoretical formalism and advanced experimental methodology. It is usually performed in three dimensions in longitudinally co-moving reference frame of the pair, to maximize the amount of information obtained about the source. Similar three-dimensional analysis has also been performed for charged and neutral kaons. In contrast analyses for all other pair types -- either for pairs of identical protons, or for pairs of non-identical particles are only one-dimensional. The abundance of heavier particles is lower than for pions, while the theoretical formalism for such correlations, as well as practical experimental methodology is limited to one-dimensional case up to now. This work identifies specific challenges arising in the analysis of three-dimensional proton-proton femtoscopy and proposes a theoretical formalism to analyze them, as well as specific experimental methods to extract and fit correlations. A test and validation of these methods is performed on correlations simulated in the LHyquid+Therminator2 model, and demonstrates the feasibility of such analysis. The caveats in the theoretical interpretation of the results are discussed, by comparing the fitted values of femtscopic radii with the true characteristics of the source extracted directly from the model. 

\end{abstract}

\pacs{25.75.-q, 25.75.Dw, 25.75.Ld}

\maketitle

%\tableofcontents

%\input{introduction}
\section{Introduction}
\label{sec:intro}

The heavy-ion collision research program aims to study in detail the phase diagram of the strongly interacting matter. Several major research facilities worldwide, such as the Large Hadron Collider (LHC) and the Super Proton Synchrotron (SPS) at CERN, Relativistic Heavy-Ion Collider (RHIC) at Brookhaven National Laboratory, as well FAIR accelerator complex at GSI aim to study specific regions of this diagram~\cite{STAR:2005gfr,ALICE:2014sbx,NA49:2008ofa,HADES:2022gdr}. At maximum collision energies reached at LHC and RHIC the quark-gluon plasma phase at small baryochemical potential is investigated. At lower energies, achieved at SPS, the Beam Energy Scan programme at RHIC as well as in the current and planned experiments at FAIR matter is investigated at lower temperatures and larger baryon densities. This region is of particular interest, as it allows to investigate conditions which are expected to occur in cores of neutron stars, as well as in neutron star mergers. Baryonic degrees of freedom are of particular importance in these studies.

One of the techniques used to investigate heavy-ion collisions is "femtoscopy", which utilizes two-particle correlations measured as a function of the relative momentum of the pair to study the spatio-temporal extent of the particle emitting sources, as well as the reaction dynamics. Recent advances in the technique allow it to be used also in the investigation of the strong interaction potential between different types of particles, which is especially useful for baryon pairs~\cite{STAR:2005rpl,ALICE:2019igo,Acharya:2018gyz,Acharya:2019yvb,Acharya:2019sms,Acharya:2019kqn,Acharya:2019bsa,Acharya:2020asf}. Femtoscopy originates in the discovery of increased production of pairs of identical pions at low relative momentum by Goldhaber, Goldhaber, Lee and Pais~\cite{Goldhaber:1960sf} and shares key similarities in mathematical formalism to the Hanbury-Brown and Twiss (HBT) method in astronomy~\cite{HanburyBrown:1954amm,HanburyBrown:1956bqd}. The (anti-)symmetrization of the pair wave function for identical particles is the crucial effect here. The method has been further developed into a standard measurement in heavy-ion collisions by Kopylov and Podgoretsky~\cite{Kopylov:1972qw,Kopylov:1973qq,Kopylov:1974th}. One of their crucial contributions was the understanding and incorporating into the formalism of the Final State Interactions (FSI) between the two particles in the pair. 

Femtoscopic investigations in high energy heavy-ion collisions naturally focus on measurements for pairs of identical charged pions, for several reasons. Firstly large multiplicities of pions produced per event allow for statistically precise measurements. The FSI for pions are the Coulomb and strong interactions. However, the strong interaction is relatively weak, and Coulomb interaction is limited to low relative momenta, so both can be treated as a small correction to the dominating effect of pair-wave symmetrization (often referred to as "Quantum Statistics" effect). For this scenario, the methodology is very well known and used extensively for several decades~\cite{Lisa:2005dd}. In particular full three-dimensional analysis is possible, well understood, and performed by practically all experiments doing femtoscopic analysis. Similarly, three-dimensional analysis for charged kaons is also possible, while for neutral kaons, where strong interaction becomes non-negligible, important complications in the experimental methodology appear~\cite{ALICE:2017iga}. In contrast, for all other pair types, including pairs of baryons, due to the limited particle multiplicities, as well as more complicated FSI, the analysis is performed only in one dimension, yielding a limited information about the source size and dynamics. Complicated FSI also requires numerical fitting procedures, which are less well tested, when compared to the analytical fitting formulas for identical pions.

Recently a full three-dimensional analysis for pairs of protons is considered for several reasons. Firstly, the study of the source dynamics in the baryonic sector is of particular interest, in order to test if any differences between mesons and baryons can be found. One-dimensional analysis is not sufficient to answer this question. The three-dimensional analysis of proton correlations at lower energies is considered as one of the important signatures of the nature of the phase transition in the QCD phase diagram. Secondly, for collisions at lower energies in RHIC BES and FAIR protons are most abundant particles, while for top LHC energies they are also produced in large numbers. In both cases, especially for the CBM experiment at FAIR as well as the upgraded ALICE experiment at LHC, the statistics should be sufficient for full three-dimensional analysis. Thirdly, developments in experimental and theoretical methodology, as well as a steady increase of available computational resources means, that numerical fitting methods that may have been prohibitively computationally expensive in the past could be within reach now. Lastly, a theoretical formalism for the three-dimensional analysis for pairs of particles where FSI is a non-negligible or dominant part of the pair wave-function (as is the case for a pair of identical protons) has reached a stage where it can be tested.

This work presents the methodology, which is a specific combination of theoretical formalism, experimental techniques, and numerical tools which allow, for the first time, to perform a meaningful three-dimensional analysis of femtoscopic correlations for pairs of protons. Section~\ref{sec:starndard} explains the standard femtoscopic analysis scenarios and their femtoscopic formalism, with particular emphasis on the specificity of the pair of protons. It is explained how previous methods are inadequate to analyze these correlations.  Section~\ref{sec:sphharm} presents arguments that the spherical harmonics decomposition is the most suitable choice for the analysis of three-dimensional functions for protons, capturing all the important features of the correlation, while also having a manageable statistics requirements. In Section~\ref{sec:ppwave} the specifics of the proton-proton wave-function are introduced. The fitting procedure is tested on the model sample. Its preparation is explained in Section~\ref{sec:testsample}. Section~\ref{sec:numfitting} describes a specific numerical tool, the CorrFit package, that is able to combine the theoretical formalism and the spherical harmonics representation to deliver a working solution for extracting three-dimensional information from the femtoscopic correlation of protons. In Section~\ref{sec:valid} a feasibility test of the method is performed on data simulated in the Lhyquid3D+Therminator2 model. It validates that the obtained "experimental-like" fit results are consistent with model inputs. Caveats in their interpretation are discussed. Lastly, specific procedures which need to be used by experiments in order to perform, for the first time, full three-dimensional analysis of proton-proton femtoscopic correlation functions are given.

\section{Femtoscopic formalism for proton pairs}
\label{sec:twopartcf}

\subsection{Standard approaches to femtoscopic analysis}
\label{sec:starndard}

In femtoscopic analysis the correlation function, defined as a ratio of the probability of registering a pair of correlated particles $P_{1,2}$ to the product of the probabilities to register them independently $P_{1} \cdot P_{2}$ is analyzed as a function of pair relative momentum $\vec{q_{inv}}=2\vec{k^{*}}=\vec{p_{1}}-\vec{p_{2}}$\footnote{Throughout the paper the following notation is used: four-vectors are written as bold: $\mathbf{r^{*}}$, three-vectors are denoted by the arrow: $\vec{k^{*}}$, while scalars are written as normal font: $q_{out}$}. When the probability to produce a given pair of particles, at a given space-time separation $\mathbf{r^{*}}$, and a given relative momentum, expressed as an "emission function" $S(\mathbf{r^{*}}, \vec{k^{*}})$ is assumed to be independent of their interaction after the emission, described by the pair wave function $\Psi$ the correlation can be expressed as the Koonin-Pratt equation:
\begin{equation}
C(\vec{k^{*}})={{\int S(\mathbf{r^{*}}, \vec{k^{*}}) |\Psi({\mathbf{r^{*}}}, \vec{k^{*}})|^{2} d^{4}r^{*}}\over{\int S(\mathbf{r^{*}}, \vec{k^{*}})  d^{4}r^{*}}} 
\label{eq:cfkpeq}
\end{equation}
Several features of this equation are important for the analysis in this paper. The space-time separation $\mathbf{r^{*}}$ must be considered as a full four-vector, as both space and time components influence how the pair is interacting. Also particle emission probability in most models depends on time, so the emission probability for a pair will depend non-trivially on their respective emission times. Femtoscopy relies on the precise description of the pair interaction, which can only be defined if both particles are identified. This means that their masses are fixed, and the relative momentum $\vec{k^{*}}$ has only three independent components. The description of the interaction between the two particles in the pair is contained in their wave-function $\Psi$. It is naturally expressed in the pair rest frame (PRF). Since other reference frames will be discussed in this work, a convention is used in which all values in PRF are always denoted by an asterisk ($^{*}$). Importantly, the most general form of the correlation function, defined by Eq.~\eqref{eq:cfkpeq} is a three-dimensional object. 

In standard analysis of the femtoscopic correlation function one of the two strategies are usually employed. The first is used for pairs of identical charged pions, and identical charged kaons. It uses a particular reference frame, as well as a specific decomposition of the relative momentum and other vectors. Pion femtoscopy is used as a probe of system collectivity, which naturally arises in hydrodynamic models. There, particle emission proceeds from a given fluid element. In order to approximate the reference frame of this element, the femtoscopic analysis is performed in the Longitudinally Co-Moving System (LCMS), where the total momentum of the pair in the direction of the beam (the "longitudinal" direction - or $z$) vanishes: $p_{1,z} = -p_{2,z}$. All vectors in this frame are then expressed in the Bertsch-Pratt decomposition, where, in addition to the longitudinal direction, the "outward" direction is defined as a direction of the total transverse momentum of the pair, and the third direction, perpendicular to the other two is referred to as "sidewards". The source emission function is expressed in this choice of variables:
\begin{equation}
S(\mathbf{r^{*}}, \vec{k^{*}}) = \exp \left ({{{-r_{out}^{2}}\over{4R_{out}^2}}+{{-r_{side}^{2}}\over{4R_{side}^2}}+{{-r_{long}^{2}}\over{4R_{long}^2}}} \right ) .
\label{eq:srclcms}
\end{equation}
Important features of this form are as follows. The emission is assumed to be a three-dimensional ellipsoid with Gaussian density profile, with three different widths: $R_{out}$, $R_{side}$, and $R_{long}$ in three directions, defined in LCMS. This assumption is again motivated by theoretical models of the heavy-ion collisions, and so far has been successful in describing experimental data in such collisions~\cite{ALICE:2015tra,ALICE:2017gxt}. The independence of the function on relative momentum is called the smoothness approximation, and also has been validated experimentally. The emission function $S$ depends on a "relative separation" $\mathbf{r^{*}}$ of the two particles. Note that the emission time difference does not appear directly in the function. The significance of this fact is discussed in detail in Sec.~\ref{sec:routlcmsorigin}. For the specific case of identical particles their relative separation is a convolution of two identical single-particle emission functions. If they are Gaussians, their convolution is also a Gaussian, but with a width multiplied by $\sqrt{2}$. Therefore the formulation in Eq.~\eqref{eq:srclcms}, with the factor $4$ in the denominator ensures, that the $R$ parameters have a straightforward interpretation of single-particle source sizes in LCMS, which is of direct interest in comparisons with models, discussed in Section~\ref{sec:valid}. 

For a pair of identical charged pions, the wave function $\Psi$ must be properly symmetrized, and describe their FSI, which includes both Coulomb and Strong interaction. However, in this specific case the strong interaction is relatively weak, and can be neglected. The influence of the Coulomb interaction can be approximated to be independent of the effects of quantum statistics, and treated separately. Then the pair wave function has a straightforward form:
\begin{equation}
|\Psi_{\pi\pi,QS}|^{2} = 1 + \cos(2\vec{k^{*}}\vec{r^{*}}) .
\label{eq:pipipqssi}
\end{equation}
In the argument of the cosine the fourth component of the relative separation would be multiplied by the energy difference, which is zero by construction in PRF for identical particles. Therefore relative separation simplifies to a three-vector in this case. Note, that unlike the definition of the source function in Eq.~\eqref{eq:srclcms}, the relative separation is expressed in PRF here. Entering $S$ defined by Eq.~\eqref{eq:srclcms}, and $\Psi$ defined by Eq.~\eqref{eq:pipipqssi} into Eq.~\eqref{eq:cfkpeq} yields the following form for the correlation function:
\begin{equation}
C(\vec{q}_{inv}) = 1 + \lambda \exp \left ( -R_{out}^{2}q_{out}^2 -R_{side}^{2}q_{side}^{2} - R_{long}^{2}q_{long}^{2}\right )    
\label{eq:pipicfqs}
\end{equation}
It is a three-dimensional function defined in LCMS. The inclusion of the Coulomb FSI is done using a Bowler-Sinyukov approximation~\cite{Sinyukov:1998fc}, where the Coulomb is introduced as a specific multiplicative factor, that weakly depends on the general size of the size, and can be treated as a correction. Such formula can then be analytically fitted to the experimental correlation function and directly yield the single particle source parameters $R$, in three directions, in LCMS. This methodology has been consistently used universally in all heavy-ion experiments to extract with great precision source parameters for pions in many collision energies, as a function of event centrality, pair transverse momentum and event plane orientation. The specific conditions, that must be met in order for this strategy to be feasible are as follows. Firstly, the number of pairs must be sufficient to provide statistically significant sample in every bin of a three-dimensional histogram of experimentally measured correlation function $C(\vec{q}_{inv})$. Secondly the FSI for a given pair must be small enough, so that they can either be neglected (e.g. strong interaction for pions), or treated as a correction (e.g. Coulomb for pions) to the dominant effect of quantum statistics. Thirdly, only one very specific choice of $S$ - a three-dimensional ellipsoid with Gaussian density profile, combined with a relatively simple form for $\Psi$ in Eq.~\eqref{eq:pipipqssi} allows the integral in Eq.~\eqref{eq:cfkpeq} to be performed analytically and yield a simple formula in Eq.\eqref{eq:pipicfqs}. In all other cases the integral is considerably more complicated and usually requires numerical integration. Note that none of the specific requirements mentioned above are fulfilled for the pair of identical protons. Therefore, this strategy is not applicable for the analysis for this type of pair.

The second strategy is used when some or all requirements needed in the first one are not met. For example, for most particle types other than pions or kaons, statistics is often too limited for full three-dimensional analysis. For pairs of non-identical particles, or for pairs of protons, the effects of FSI cannot be neglected, and are rather the dominant part of the wave-function (see also Sec.~\ref{sec:ppwave}). In that case the correlation function is constructed as a function of $k^{*}$ magnitude only. This amounts to the following modification of Eq.~\eqref{eq:cfkpeq}:
\begin{equation}
C({k^{*}})={{\int S(\mathbf{r^{*}}, \vec{k^{*}}) |\Psi({\mathbf{r^{*}}}, \vec{k^{*}})|^{2} {k^{*}}^{2} d\cos\theta_{k}d\varphi_{k} d^{4}r^{*}}\over{\int S(\mathbf{r^{*}}, \vec{k^{*}}) {k^{*}}^{2} d\cos\theta_{k}d\varphi_{k} d^{4}r^{*}}}
\label{eq:cfkpeqinv}
\end{equation}
The integration, both in numerator and denominator is now additionally performed over full angular space in relative momentum. So the resulting correlation function could be understood as directionally averaged. The crucial limitation of this method arises precisely from this averaging. Such correlation function looses sensitivity to the differences in source radii in the three Betsch-Pratt components. That is why the source function assumption of Eq.~\eqref{eq:srclcms} must also be simplified to:
\begin{equation}
S(\mathbf{r^{*}}, \vec{k^{*}}) = \exp \left (-{{{{r^{*}_{out}}^{2}+{r^{*}_{side}}^{2}+{r^{*}_{long}}^{2}}\over{4R_{inv}^2}}} \right )
\label{eq:srclcms1d}
\end{equation}
Two important differences appear - in this strategy the radius $R_{inv}$ is usually defined in PRF rather than in LCMS, and it is the only parameter of the source function. Note however, that the source function itself, as well as the integration in Eq.~\eqref{eq:cfkpeqinv} over the space-time separation remain four-dimensional. 

The other important feature of this strategy is the fact, that for cases of interest the simplified form of the wave function in Eq.~\eqref{eq:pipipqssi} is no longer sufficient. A full form of $\Psi$, usually containing the FSI contributions must be used. It is discussed it in detail in Sec.~\ref{sec:ppwave}. Then, it is not possible to perform the integration in Eq.~\eqref{eq:cfkpeqinv} analytically. Instead various numerical methods can be employed. A common strategy is as follows. From events generated by Monte-Carlo models, combine the particles into pairs, and calculate their relative momentum. Store it in a histogram, which is the denominator in Eq.~\eqref{eq:cfkpeqinv}. Calculate the pair wave function modulus squared, and store it in another histogram at the same relative momentum, constituting the numerator of Eq.~\eqref{eq:cfkpeqinv}. Note that for this calculation, the space-time emission points of each particle must be either taken from the model, or generated, for example according to the Eq.~\eqref{eq:srclcms1d}. Dividing the two histograms one obtains the correlation function $C(k^{*})$. This procedure is simply the Monte-Carlo calculation of both integrals in Eq.~\eqref{eq:cfkpeqinv}.

The obvious limitation of this strategy is the fact that it provides only a single directionally-averaged estimation of the source size, not the full three-dimensional information. Also the calculation of the correlation function (which is an essential part of the procedure of fitting the experimental correlation function) is a potentially numerically intensive procedure of Monte-Carlo integration. But the procedure also has its advantages. It requires significantly less statistics to provide a good estimate of the source size. It can use a pair wave-function even it is rather complicated, which means that it can be employed for any pair - identical or non-identical. Also, since the integration in Eq.~\eqref{eq:cfkpeqinv} is fully numerical anyway it means that other forms of the source function $S$, not only Gaussian ellipsoids, can be easily explored. 

The purpose of this work is the formulation and a full model validation of another strategy to analyze the femtoscopic correlation functions, which combines the advantages of the two strategies mentioned above, while avoiding most of their limitations, and still remaining computationally viable. The procedure remains fully flexible, since it still relies on the numerical integration. But instead of using a three-dimensional Cartesian representation of the correlation function, it relies on its spherical harmonics representation. Since only three components of this representation contain most of the relevant information, this representation requires vastly smaller number of pairs to produce a statistically significant dataset. It is also more straightforward to visualize a full three-dimensional structure of the correlation. And most importantly this procedure allows to extract the full three-dimensional information about the source sizes. As a result this strategy is ideal for the full three-dimensional analysis for pairs of identical protons, and it is used for this purpose in this work. The following sections describes all the concepts and mathematical tools necessary for the execution of this strategy.

\subsection{Spherical Harmonics representation}
\label{sec:sphharm}

A three-dimensional femtoscopic correlation function can be represented as a series of one-dimensional histograms, corresponding to a spherical harmonics (SH) decomposition of the function:
\begin{equation}
    C_{l}^{m}(k^{*}) = \int C(\vec{k^{*}}) Y_{l}^{m}(\theta_{k}, \varphi_{k}) d\cos(\theta_{k}) d\varphi_{k} ,
    \label{eq:cylmdef}
\end{equation}
where $Y_{l}^{m}(\theta_{k}, \varphi_{k})$ are complex spherical harmonics functions, with $l$ an integer running from 0 to infinity, and $m$ an integer in the range of $-l$ to $l$. The "longitudinal" Bertsch-Pratt direction is the $z$ axis, vs which the $\theta_{k}$ angle is determined, the "out" is associated with $x$, and the "side" with $y$ in the transverse plane where the $\varphi_{k}$ is defined. Generally, the full behaviour of an arbitrary function can only be reproduced if sufficient, potentially large number of these components is used. However, femtoscopic correlation functions have particular features, which result in a significant reduction of relevant components. Since $C$ is real, even-$l$ imaginary components vanish. Identical particles introduce particle-interchange symmetry which causes all odd-$l$ components to vanish. For boost-invariant, azimuthally symmetric source odd-$m$ components for even-$l$ terms also vanish. Finally, as the $l$ number increases, the components correspond to more detailed information about the angular distribution of the correlation. However, for all realistic femtoscopic correlation functions, including the ones for identical protons, the correlation function is expected to smoothly change in angular space. This means that higher-$l$ components will either vanish, or not carry additional information relevant for extracting source radii. Consequently, only three components of the spherical composition are relevant for the analysis - the $C_{0}^{0}$ is sensitive to overall directionally-averaged size of the source, $C_{2}^{0}$ deviates from zero when longitudinal radius $R_{long}$ differs from the transverse ones, and the $C_{2}^{2}$ component reflects the difference between $R_{out}$ and $R_{side}$\cite{Kisiel:2009eh}. As a result spherical harmonics decomposition is found to be extremely efficient in representing all relevant features of the correlation function - instead of having a full three-dimensional matrix of bins, there are only three one-dimensional histograms, with significantly smaller statistics requirements~\cite{Kisiel:2009iw}.

In order to fully utilize the efficiency of the SH representation explained above the way in which the SH components are actually determined is important. In particular if, in order to perform the integration in~\eqref{eq:cylmdef}, 3D histogram histogram (in Cartesian $x,y,z$ or angular $r,\theta,\varphi$ coordinates) is used, the advantage is nullified. That is why a special method was developed~\cite{Kisiel:2009iw}, where the numerator and the denominator of the correlation function is constructed directly in spherical harmonics, without the need for any intermediary 3D representation. The correlation function is also directly computed in SH, preserving the advantages of this representation. The usage of spherical harmonics, in the specific implementation mentioned above, without any intermediate 3D stage is used in this work.

\subsection{The proton-proton wave function}
\label{sec:ppwave}

\begin{figure}[!h]
\centering\includegraphics[width=.999\linewidth]{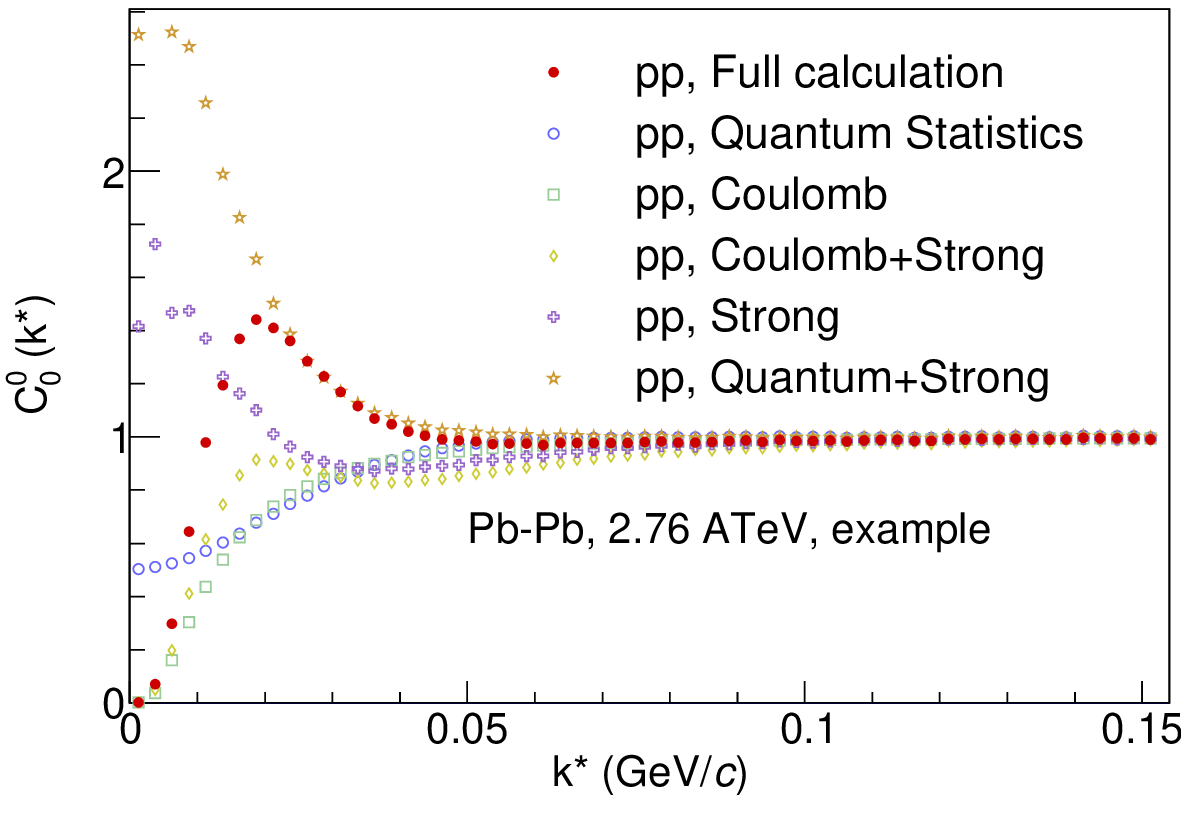}
\caption{Proton-proton femtoscopic correlation function for an example centrality and transverse momentum range typical for Pb-Pb collisions at LHC energies. Full circles show full calculation, while open symbols correspond to selected combinations of partial contributions of quantum statistics and FSI effects.}
\label{fig:ppcorrexample}
\end{figure}

The proton-proton system, being composed of two fermions, has two contributions to the wave function: in the singlet (S) and triplet (T) state. The wave function for the S component is symmetrized, and is counted with weight $1/4$, while the T component is anti-symmetrized and counted with $3/4$ weight. So pure QS function would exhibit a negative correlation approaching 0.5 at zero $q_{inv}$. In addition protons are charged hadrons, therefore are subject to both Coulomb and strong FSI. Their combination is expressed as~\cite{Lednicky:2005tb, Lednicky:1981su, Lednicky:2003mq}:
\begin{equation}
    \Psi_{-k^{*}}(\mathbf{r^{*}})=\sqrt{A_c(\eta)} \left [ e^{i \mathbf{k^{*} \mathbf{r^{*}}} }F(-i\eta,1,i\xi) +f_{c}\tilde{G}(\rho, \eta)/r^{*} \right ]
\label{eq:scfsi}
\end{equation}
\noindent where $\rho = \mathbf{k^{*}\mathbf{r^{*}}}$, $\xi = \mathbf{k^{*}}\mathbf{r^{*}}+k^{*}r^{*}$, $\eta=(k^{*}a_C)^{-1}$ and $a_C$ is the Bohr radius of the pair, equal to $57.6$~fm for a proton pair. $A_C$ is the Gamow factor, $F$ is a confluent hypergeometric function and $\tilde{G}$ is a combination or regular and singular s-wave Coulomb functions. $f_{C}$ is the strong scattering amplitude modified by the Coulomb interaction. Applying Eq.~\eqref{eq:scfsi} for identical fermions requires the calculation of the singlet and triplet components separately, proper symmetrization the singlet and anti-symmetrization the triplet component and then calculation of their sum with relevant weights. For large source sizes, like the ones considered int this work, the $f_C$ can be expressed in the "effective range" approximation as:  
\begin{equation}
f_C(k^{*}) = \left [ {{1} \over {f_0}}+{{1}\over{2}}d_0 {k^{*}}^{2}-{{2}\over{a_C}}h(k^{*}a_C)-ik^{*}A_C(k^{*})  \right]^{-1}    
\label{eq:fcpp}
\end{equation}
where $h$ is related to a digamma function~\cite{Lednicky:2005tb}. The scattering length $f_0$ and the effective radius $d_0$ are parameters characterizing the strong interaction, which differ for singlet and triplet configuration. In this work the following values are used: $f_0^S=7.77$~fm, $d_0^S=2.77$~fm, $f_0^T=-5.4$~fm, $d_0^T=1.7$~fm. 

With the proton-proton wave function defined above the final form of the proton-proton correlation function can be numerically simulated. In Fig.~\ref{fig:ppcorrexample} several example calculations are shown for a sample representative of Pb-Pb collisions at LHC energies. The full calculation, including effects of Quantum Statistics, as well as strong and Coulomb FSI shows a characteristic shape. It goes to zero at vanishing $k^{*}$ due to Coulomb repulsion. Even though net QS effect is negative, the addition of strong interaction is able to overcome this and produces a distinct peak around $k^{*}=20$~MeV/$c$. In summary, the exact shape of the proton-proton correlation is a non-trivial combination of all three components - QS, strong and Coulomb. They do not factorize, and none of them can be neglected or treated as a small correction. As a result fitting of the proton-proton correlation must always include all effects in the calculation, and is performed numerically. 

In this work the formulation of the proton-proton wave-function described above is used self-consistently to first simulate the correlation function from the model directly, and then to perform the fitting procedure. More sophisticated descriptions of the proton-proton wave-function exist~\cite{ALICE:2018ysd}, and this subject is an area of intensive theoretical developments. They are not needed in this work, as the formalism is realistic enough for the model considerations, relatively efficient numerically, and the critical part is the self-consistency between model correlation function calculation and fitting. Nevertheless, in fitting real experimental data it might be advisable to use more advanced form of the wave-function. The fitting strategy is fully compatible with such procedure. One needs to simply replace the part of the numerical code that calculates the wave function with the code implementing the more advanced theoretical formulation. The rest of the procedure remains unchanged and all the conclusions of this work remain valid.

\section{Preparation of the test sample}
\label{sec:testsample}

The aim of this section is to perform a validation of the procedure of three-dimensional fitting of the full proton-proton correlation function on specific model input. For this exercise a sample of events used is simulated in a combination of LHyquid3D hydrodynamic model and Therminator2 hadronization, resonance propagation and decay code. This model provides momenta and freeze-out coordinates of all produced particles, which is crucial for the calculation of the correlation function according to Eq.~\eqref{eq:cfkpeq}. The details of the model have been discussed extensively in~\cite{Bozek:2011ua, Kisiel:2014upa, Chojnacki:2011hb}. It has been tuned to reproduce Pb-Pb collisions at $\sqrt{s_{NN}}=2.76$~TeV. In particular it is able to describe in detail the space-time characteristics of the pion source measured by ALICE at the LHC~\cite{ALICE:2015tra}. This work is not focused on the features of the model, but rather treats it as an useful and realistic simulation of the environment in such collisions. 

The study in this work uses selected event samples generated with the LHyquid3D+Therminator2 code tuned to Pb-Pb collisions divided into samples corresponding to centralities: $0-10\%$, $10-20\%$, $20-30\%$, $30-40\%$, and $40-50\%$. Protons from these events are combined into pairs, and such pairs are divided into samples based on the total transverse momentum of the pair $k_{T}=|\vec{p}_{T,1}+\vec{p}_{T,2}|/2$. Four ranges are used: $0.4-0.8$~GeV/$c$, $0.8-1.0$~GeV/$c$, $1.0-1.2$~GeV/$c$, and $1.2-1.6$~GeV/$c$. Such choice covers the typical acceptance of the ALICE experiment at the LHC, provides enough data points to see the evolution of the correlation function with pair momentum and produces 20 independent correlation functions with differing features, which should be sufficient for thorough validation of the method. 

\begin{figure}[!h]
\centering\includegraphics[width=.999\linewidth]{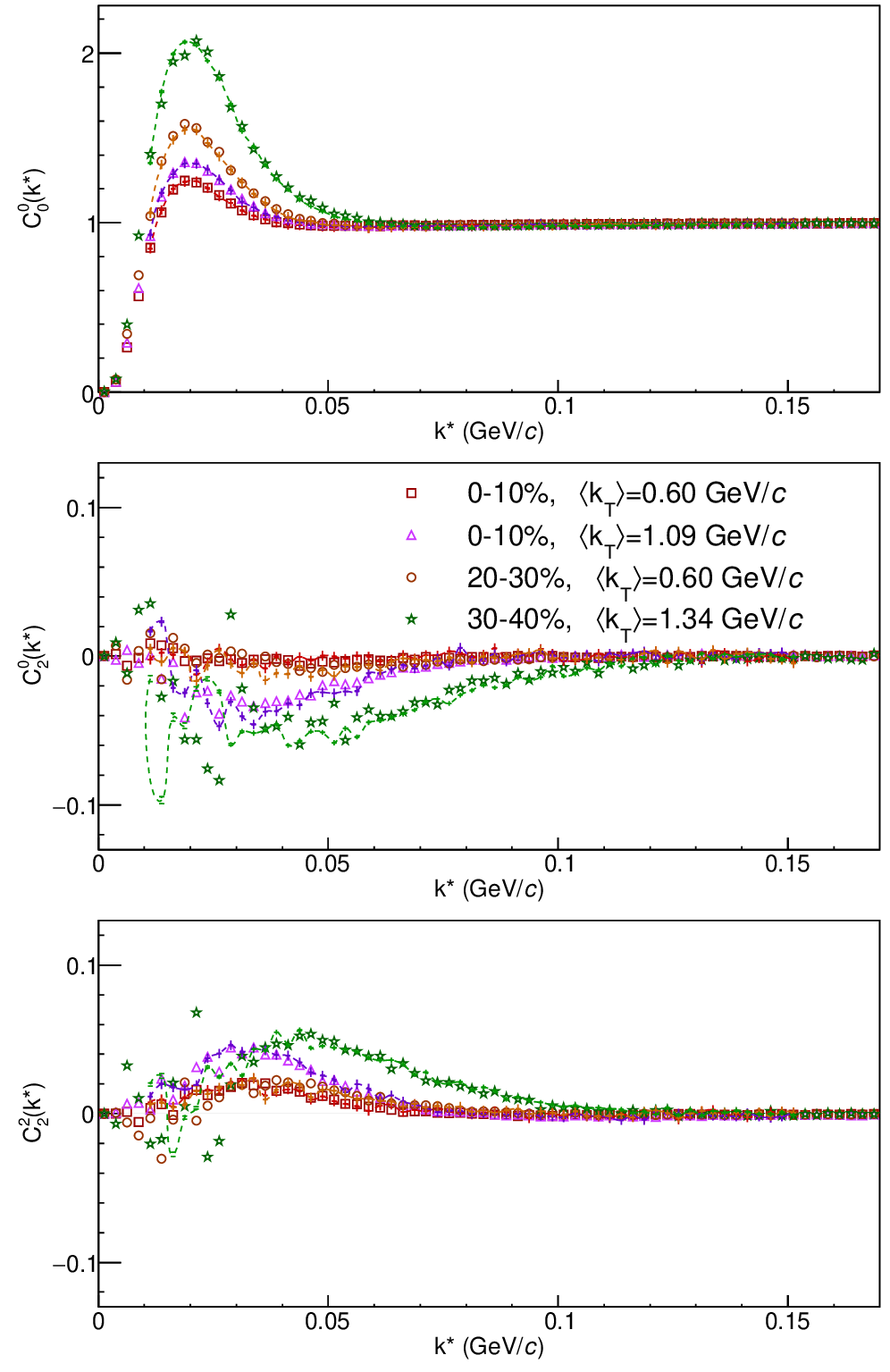}
\caption{Proton-proton correlation functions for example ranges of centrality and $k_{T}$. Points show correlation functions calculated directly from the model. Lines represent fits with the procedure described in this work.}
\label{fig:ppfunsample}
\end{figure}

The correlation function is then constructed in the following manner. For each pair of protons its relative momentum $k^{*}$ is calculated. Then this pair is added to the denominator of the correlation with weight 1.0, directly in spherical harmonics. Then, the full modulus squared of the wave function is calculated for this pair, according to the procedure described in Sec.~\ref{sec:ppwave}. Then the pair is added to the numerator of the correlation function with this weight, again directly in spherical harmonics. Finally the numerator is divided by the denominator to produce the correlation function. The procedure to perform this operation fully in spherical harmonics representation (without any need for intermediary three-dimensional histograms) is described in~\cite{Kisiel:2009iw}. In this way 20 correlation functions for all centrality/$k_{T}$ ranges are calculated and then analyzed in further sections. Example correlation functions are shown in Fig.~\ref{fig:ppfunsample}. The $C_0^0$ component for all functions exhibit the characteristic peak. Its height differs significantly between samples, which reflects the expected difference in femtoscopic radii. The other two components $C_2^0$ and $C_2^2$ show statistically significant deviations from unity, which is expected if the out, side and long radii differ for a particular sample. The aim of the fitting procedure developed in this work is to describe these deviations and use them to extract the values of all three radii separately. 

\begin{figure}[!h]
\centering\includegraphics[width=.999\linewidth]{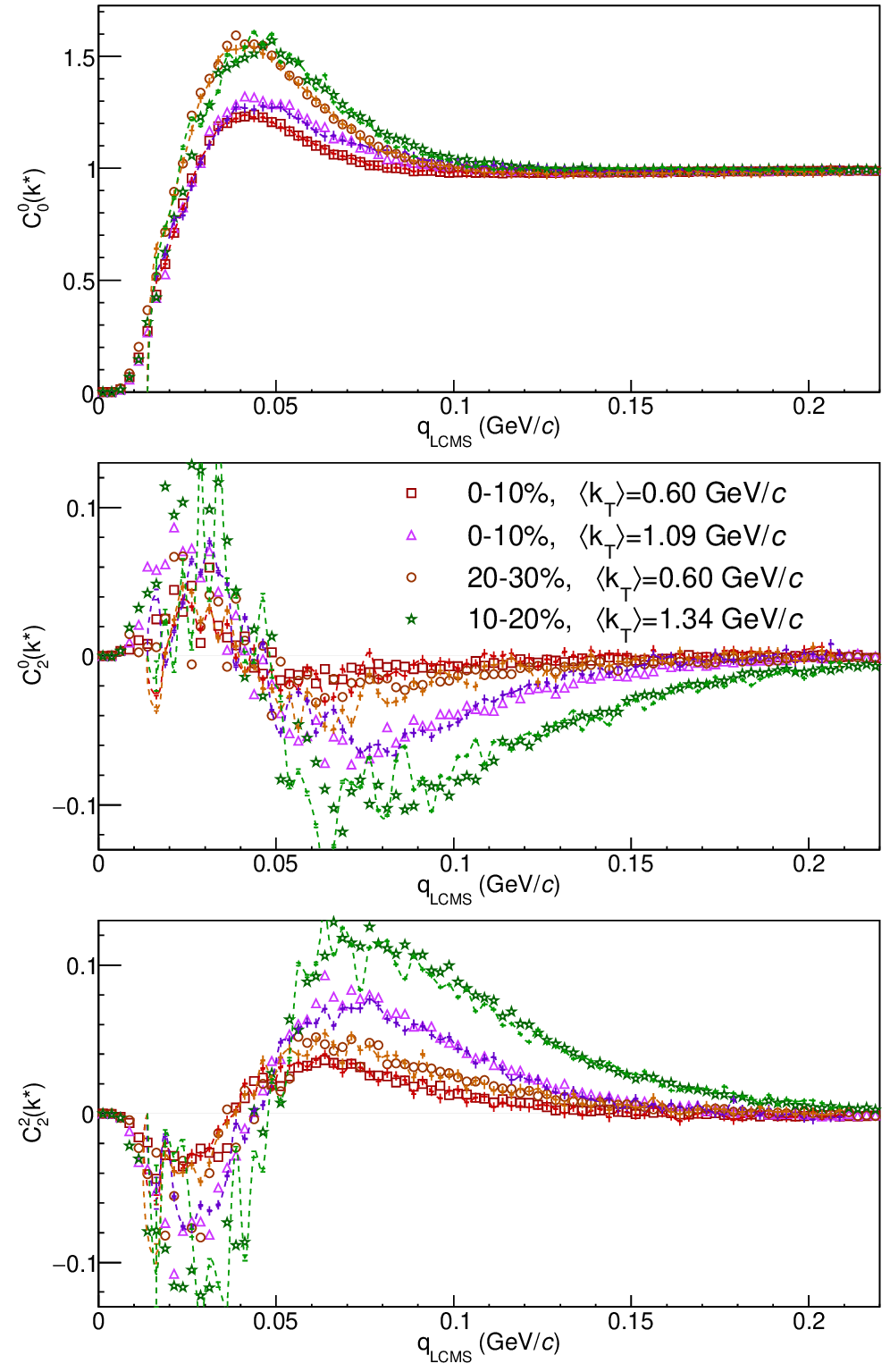}
\caption{Proton-proton correlation functions for example ranges of centrality and $k_{T}$, calculated as a function of the relative momentum in the LCMS frame. Points show correlation functions calculated directly from the model. Lines represent fits with the procedure described in this work.}
\label{fig:ppfunsamplelcms}
\end{figure}

The same procedure has been repeated, but the histograms have been filled with the relative momentum calculated in LCMS, instead of PRF. Examples are shown in Fig.~\ref{fig:ppfunsamplelcms}. Following the procedure standard in identical pion femtoscopy the variable plotted is $q_{LCMS}$, the full relative momentum of the pair, as opposed to $k^{*}$ which is half of $q_{inv}$, in the example above. The two representations should be complementary, which will be tested in this work. However, the SH representation in LCMS, especially for $l=2$ exhibits more non-trivial behaviour. Nevertheless also in this representation a statistically significant signal is present in all SH components.

The correlation function simulated in this way includes only the femtscopic correlation component. It does not account for important effects present in experimental data, such as momentum resolution, particle identification purity, or residual correlations. However, such effects cannot be easily modeled in an experiment-independent way. One needs to rely on the experimental collaborations to properly account for them, in a way specific to their detector, and produce corrected correlation functions suitable for fitting with the procedure described in this work. Nevertheless some of these effects, such as momentum resolution could be also implemented as a modification of the numerical procedure described later. Others, such as residual correlations, can be modeled in the LHyquid+Therminator2 code. Such study is beyond the scope of this work, and can be performed in the future. 

\section{Numerical fitting procedure}
\label{sec:numfitting}

The numerical fitting procedure uses the CorrFit package~\cite{Kisiel:2004fcn}. For this work the package was developed to allow for fitting of the correlation functions in Spherical Harmonics representation. In addition the functionality to have three independent parameters in the fit (such as the three radii) has been added. The fitting algorithm proceeds as follows. 1) The sample of pair momenta is prepared based on model events. This amounts to storing all components of momenta of both particles in the pair for a given centrality/$k_T$ sample, for a limited number of pairs. Analogous procedure can be performed in the experiment, where true experimentally measured momenta of particles are stored. Such pair sample then has the exact kinematic constraints as the pairs that are used to form the correlation function. 2) A specific functional form for the source model is selected. For the reasons explained above the three-dimensional ellipsoid with the Gaussian density profile in LCMS defined in Eq.~\eqref{eq:srclcms} is used. This model has three parameters: $R_{out}$, $R_{side}$, and $R_{long}$, corresponding to the widths of the source (radii) in three Bertsch-Pratt directions in the LCMS. 3) Ranges in which the fitting of all parameters will be performed is selected. For this study the same range is used for all radius parameters: from $1.4$~fm up to $4.25$. All resulting parameters lie in this range, confirming that the chosen range is sufficient. 4) In these ranges a grid of equidistant points is constructed, each point corresponding to a specific combination of the three parameters. 5) The "experimental" correlation function to be fitted is read by the program, determining the reference frame in which the procedure will take place (at the moment fitting of correlation functions stored as a function of $\vec{q}_{LCMS}$ or $\vec{k}^{*}$/$\vec{q}_{inv}$ is implemented). The fitting and normalization range in the relevant variable is also determined. 6) At each grid point the "theoretical" correlation function is calculated numerically in the following way. Each pair of the sample constructed in step 2) is read. For both particles the freeze-out coordinates are generated with a probability distribution defined by Eq.~\eqref{eq:srclcms}, and a "weight" $w_c=|\Psi|^{2}$ is calculated for each pair. The denominator is filled with 1.0 and the numerator is filled with $w_{c}$ for each pair at its relative momentum. At the end the "theoretical" correlation function is calculated as a ratio of numerator to denominator, with proper normalization. 7) At each grid point a $\chi^{2}$ value is calculated, by comparing the "theoretical" correlation function calculated in step 6) and the "experimental" one read in step 5). As a result a three-dimensional grid of $\chi^{2}$ values is obtained, which can be stored and examined further, to assess the quality and validity of the fit. The parameters of the grid point, for which the minimum value of $\chi^{2}$ is found is the result of the fit. The "best fit" "theoretical" correlation function is also stored. Examples of such functions are shown as lines in Figs.~\ref{fig:ppfunsample} and~\ref{fig:ppfunsamplelcms}. Similarly, the distributions of $r_{out}$, $r_{side}$ and $r_{long}$, both for LCMS and PRF, which produced the "best-fit" correlation functions are also stored.

The general algorithm described above can be further customized as necessary. The procedure is numerically intensive, so the initial fitting may be done with relatively large grid spacing and repeated in the vicinity of the minimum with finer spacing. Such procedure has been applied in this work, with the fine spacing of approx.~$0.4$~fm for each parameter. When the program was executed on a modern desktop PC a calculation for an individual grid point takes less than 2 minutes. A standard fitting grid has on the order of 1000 points. The CorrFit software allows for fully parallel calculation of each grid point, so a single fitting run on a 32-thread desktop CPU takes a few hours. The calculated functions are stored in the database, so subsequent fits which re-use these calculated datapoints can proceed much quicker, even in a matter of seconds if no new calculations are needed. In summary, the numerical fitting procedure described above has low enough computational requirements, so that it can be practically executed with relative ease, even on commodity PC hardware currently available.

The fitting based on numerical integration, described above and implemented in CorrFit has a large degree of flexibility. If needed, any new form of the source function can be implemented, not limited to Gaussian profiles, and expressed in any reference frame. The numerical code to calculate the pair wave-function can be developed to include any formulation of the FSI potentials, including the ones corresponding to recent theoretical developments. In this work, as an example of this flexibility, the correlation functions were prepared and then fitted using Quantum Statistics only, in order to investigate the importance of FSI in the procedure.

\subsection{Three-dimensional and one-dimensional fit results}
\label{sec:3dfitting}

\begin{figure}[!h]
\centering\includegraphics[width=.999\linewidth]{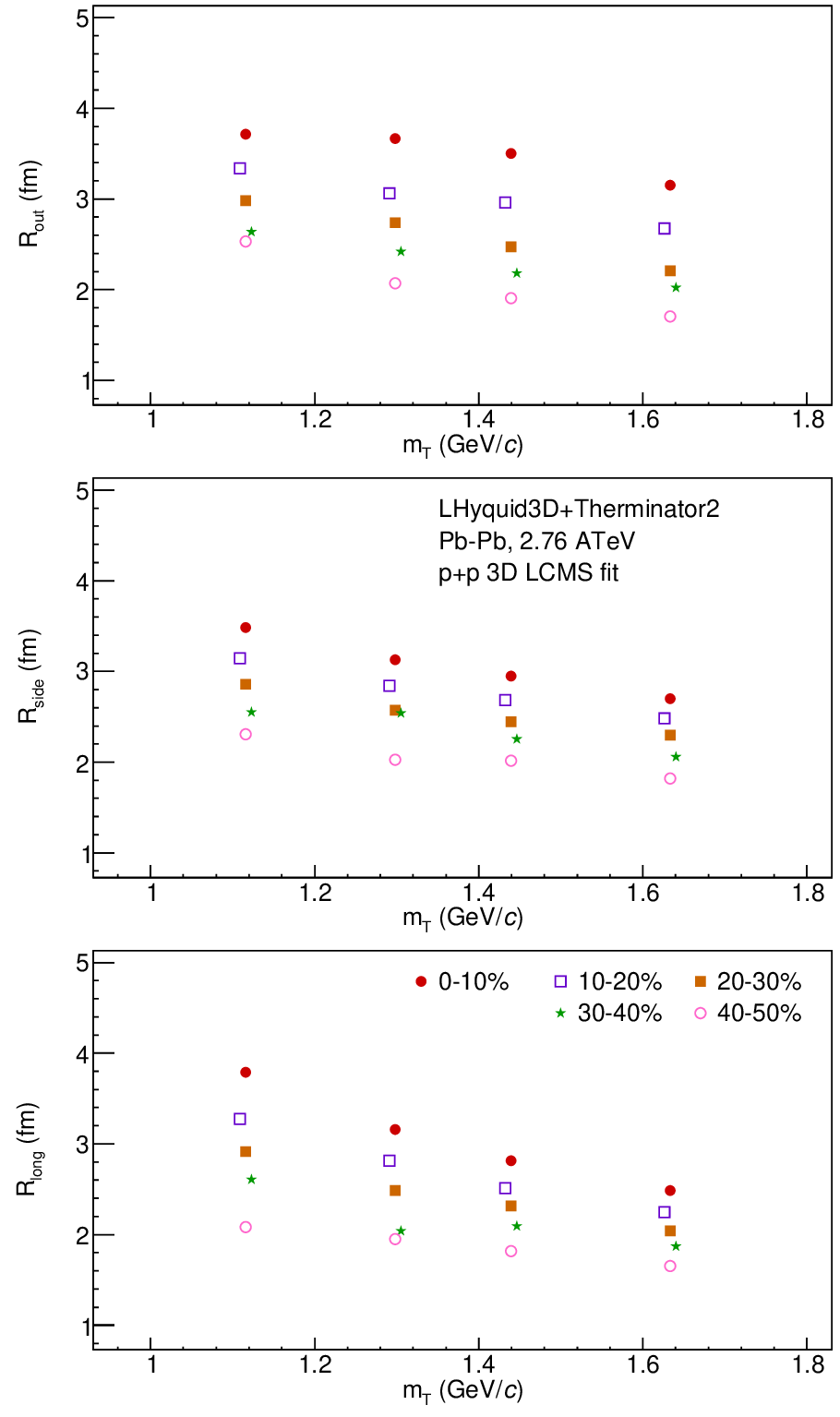}
\caption{Results of the three-dimensional LCMS fits of proton-proton correlation function simulated in Lhyquid3D+Therminator2 model for Pb-Pb collisions at $2.76$~ATeV. Values are presented versus collision centrality and pair transverse momentum. Selected points are slightly shifted in the $x$ direction for readability.}
\label{fig:fit3dlcmsradii}
\end{figure}

Example "experimental-like" correlation functions, as well as the "theoretical" "best-fit" functions are shown in Figs.~\ref{fig:ppfunsample} and~\ref{fig:ppfunsamplelcms}. The radii dependence on collision centrality and pair transverse momentum is summarized in Fig.~\ref{fig:fit3dlcmsradii}.  The radii show expected trends - they increase with increasing event multiplicity (lowering centrality) and they decrease with the growth of $k_{\rm{T}}$. This is consistent with similar simulations for all other particle types, such as charged pion and charged kaon pairs. These results are ready for comparison with experimental values, which are not yet available. The range of values obtained in the fit cover the range expected in the heavy-ion collision experiments. It is also sufficiently wide to perform the validation of the fitting method, which is described in the following sections.

\begin{figure}[!h]
\centering\includegraphics[width=.999\linewidth]{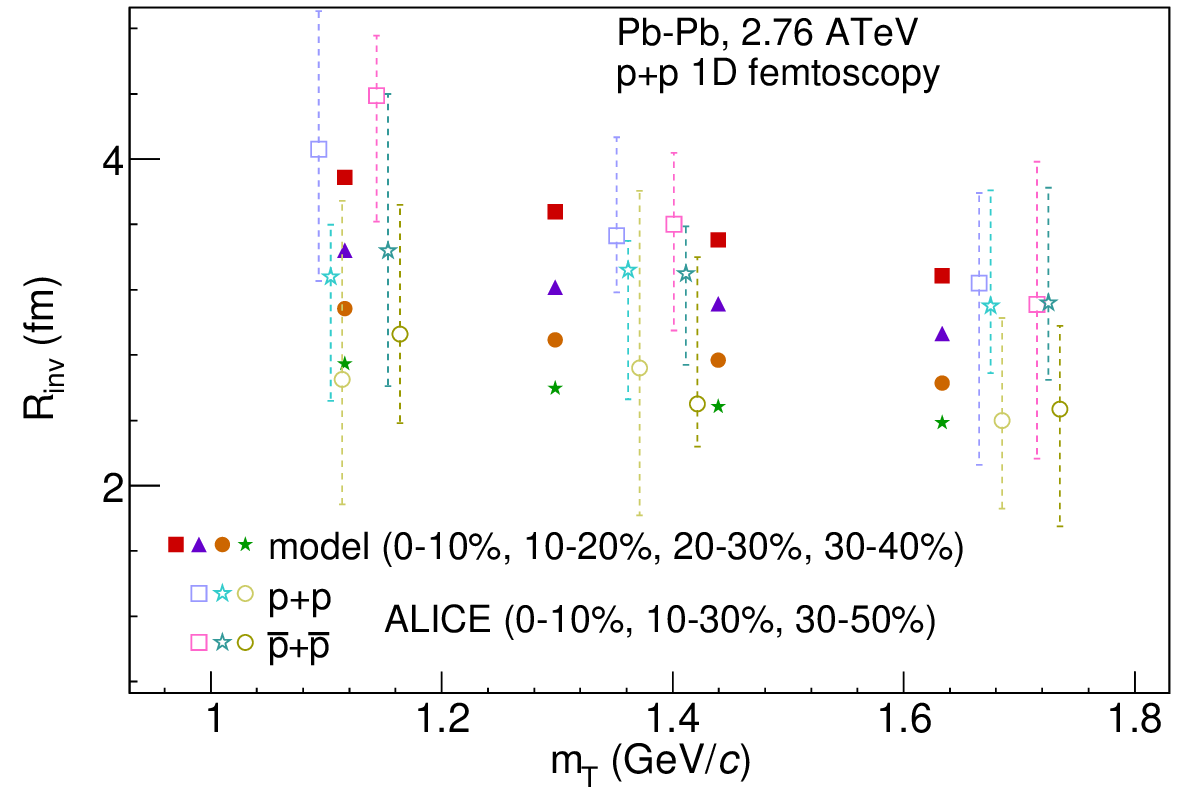}
\caption{Results of the one-dimensional PRF fits of proton-proton correlation function simulated in Lhyquid3D+Therminator2 models for Pb-Pb collisions at $2.76$~ATeV. Values are presented versus collision centrality and pair transverse momentum.}
\label{fig:fit1dprfradii}
\end{figure}

Proton femtoscopy has been studied in several heavy-ion collision experiments in the one-dimensional representation. The analyses are universally performed with the source function defined in PRF by Eq.~\eqref{eq:srclcms1d}. The numerical integration procedure has been modified to include the fitting of such functions. The results are shown in Fig.~\ref{fig:fit1dprfradii} and compared to the available data at the relevant collision system and collision energy. Expected and experimentally observed trends are visible in the fit values - growth of the radii with event multiplicity as well as a decrease with the increase of $k_{\rm{T}}$. The model values are also consistent with experimental data, confirming that the Lhyquid3D+Therminator2 model is properly tuned to reproduce the space-time characteristics of the emission process in Pb-Pb collisions at LHC energies.

\section{Validation of the fitting method}
\label{sec:valid}

\begin{figure}[!h]
\centering\includegraphics[width=.999\linewidth]{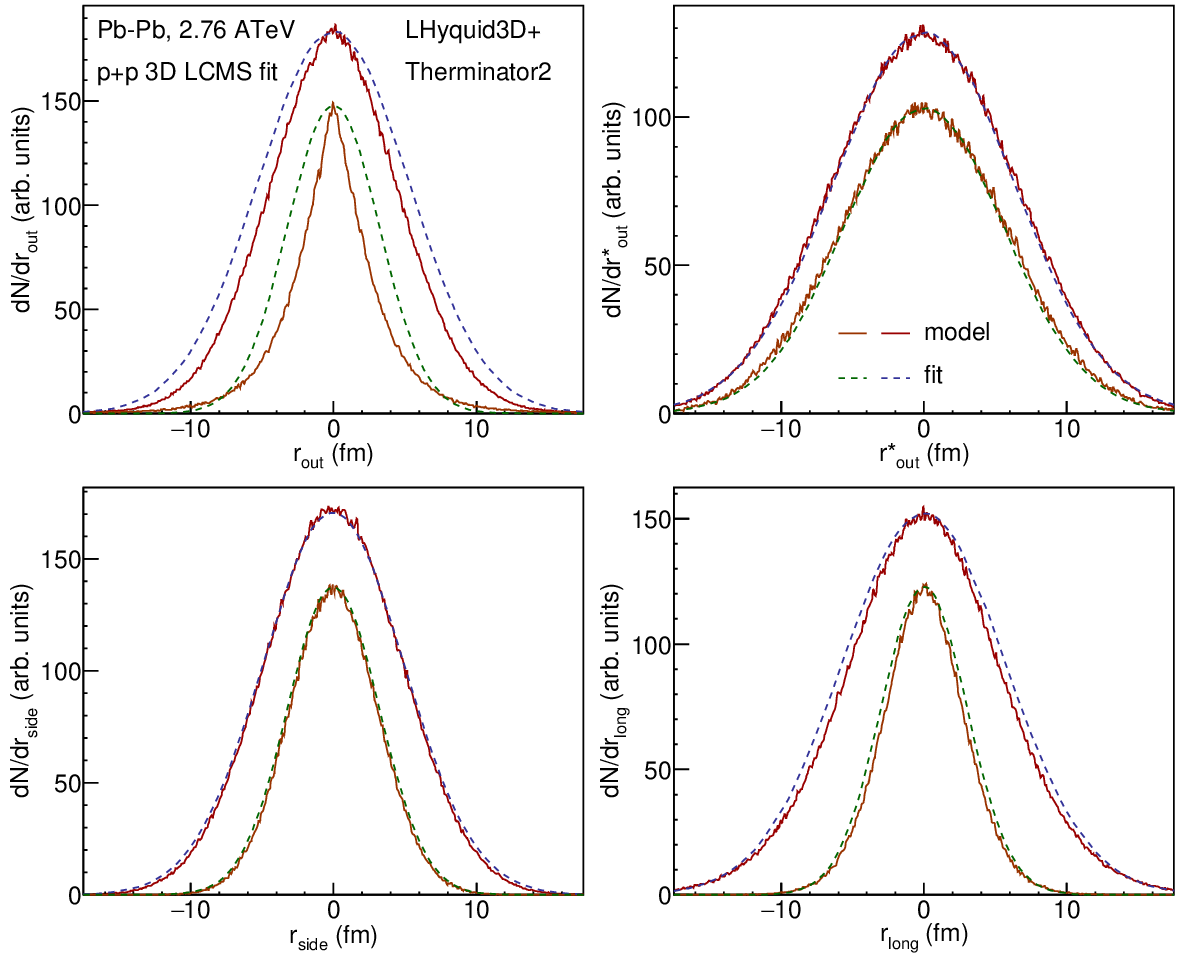}
\caption{Comparison of the input model source distributions (solid lines) to the results of the fit (dashed lines) for an example centrality and pair momentum range samples.}
\label{fig:srccomp}
\end{figure}

In the previous section it was shown that the full three-dimensional fitting with this method can be realistically performed on the simulated correlation functions, in a way that can be directly applied to experimental data. The obtained fit results should, however be validated against the model input. In Fig.~\ref{fig:srccomp} a direct comparison is shown between the input model distributions, and the source distributions, for which the best fit of the correlation function was obtained. The generation of these source distributions is an essential part of the fitting process, and they are saved in the CorrFit package as one of the cross-checks of the fit results. Note that these distributions are not fitted directly, instead only the resulting correlation functions are compared. The agreement between the input and fitted distributions is very good for the "sideward" and "longitudinal" direction at all analyzed centralities and pair momentum ranges. As it is universally done in femtoscopic analyses, the fit distributions are assumed to be Gaussian in shape. Our calculations show that the input model distributions are very similar, so this assumption of the Gaussian shape of the source is validated. Therefore, to judge the validity of the fit it is enough to compare the Gaussian widths of the input and fitted source distributions. 

The behaviour in the "outwards" direction requires more discussion. The fit reproduces well the shape and width of the distribution of $r^{*}_{out}$ in the pair rest frame, where also the Gaussian assumption is well motivated. However, there is a visible discrepancy between the input and the fit distributions in the LCMS reference frame, even though the source model that is employed in the fit is defined in this frame. The shape of the source is not precisely Gaussian, especially at large pair momentum, and the widths of the input and fit distributions differ significantly. This discrepancy in width for $r_{out}$ in LCMS is expected, but warrants a detailed explanation. As discussed in Sec.~\ref{sec:ppwave}, the interaction between particles, which determines the pair weight, and therefore also the shape of the correlation function is always calculated in PRF. Therefore, it is expected that it is the source distributions in this frame, that are reproduced by the fit. Why then, is there a discrepancy in the LCMS frame, even though the fitted source model is defined there? The values of $r^{*}_{out}$ and $r_{out}$ for a single pair are connected via the Lorentz transformation:

\begin{equation}
r^{*}_{out} = \gamma_{T} ( r_{out} - \beta_{T} \Delta t ),
\label{eq:rsovsro}
\end{equation}
and also involve the difference in emission time $\Delta t$ in LCMS. These values are calculated for every pair (with each one having a different transverse velocity $\beta_{T}$), and form the relevant distributions. The widths of these distributions should correspond to "source radii" $R_{out}$, or $R^{*}_{out}$ obtained in the fit. This simple correspondence appears to be correct for PRF, but in LCMS the role of the time difference distribution must be investigated. It must be emphasized that even though the relation between the values in two reference frames for a single pair in Eq.~\eqref{eq:rsovsro} is straightforward, there is no such simple relation between the "source radii". In particular the value of $R^{*}_{out}$ will non-trivially depend on: the width $R_{out}$ and the shape of the $r_{out}$ distribution, the width and the shape of the $\Delta t$ distribution, the possible correlation between $r_{out}$ and $\Delta t$ values, as well as the exact shape of the $\beta_{T}$ distribution for all pairs. This question is investigated in more detail in Section.`\ref{sec:routlcmsorigin}. Note, that in the fitting procedure it is possible to obtain only a single parameter for the "out" direction, so for example, there is no possibility to independently determine the width of the "space" and "time" distributions in LCMS. Only one value, which is a convolution of the two, may be obtained. Therefore the $R_{out}$ radius fitted in LCMS cannot be simply understood as the width of the $r_{out}$ distribution. Instead, in the fitting procedure described above, the width of the $\Delta t$ distribution is assumed to be zero. The fitted $R_{out}$ parameter is therefore only an "effective" radius, which must account for all the effects mentioned above, and finally produces a $r^{*}_{out}$ distribution that matches the input.

%According to Eq.~\eqref{eq:}The shape of the $r^{*}_{out}$

\begin{figure}[!h]
\centering\includegraphics[width=.999\linewidth]{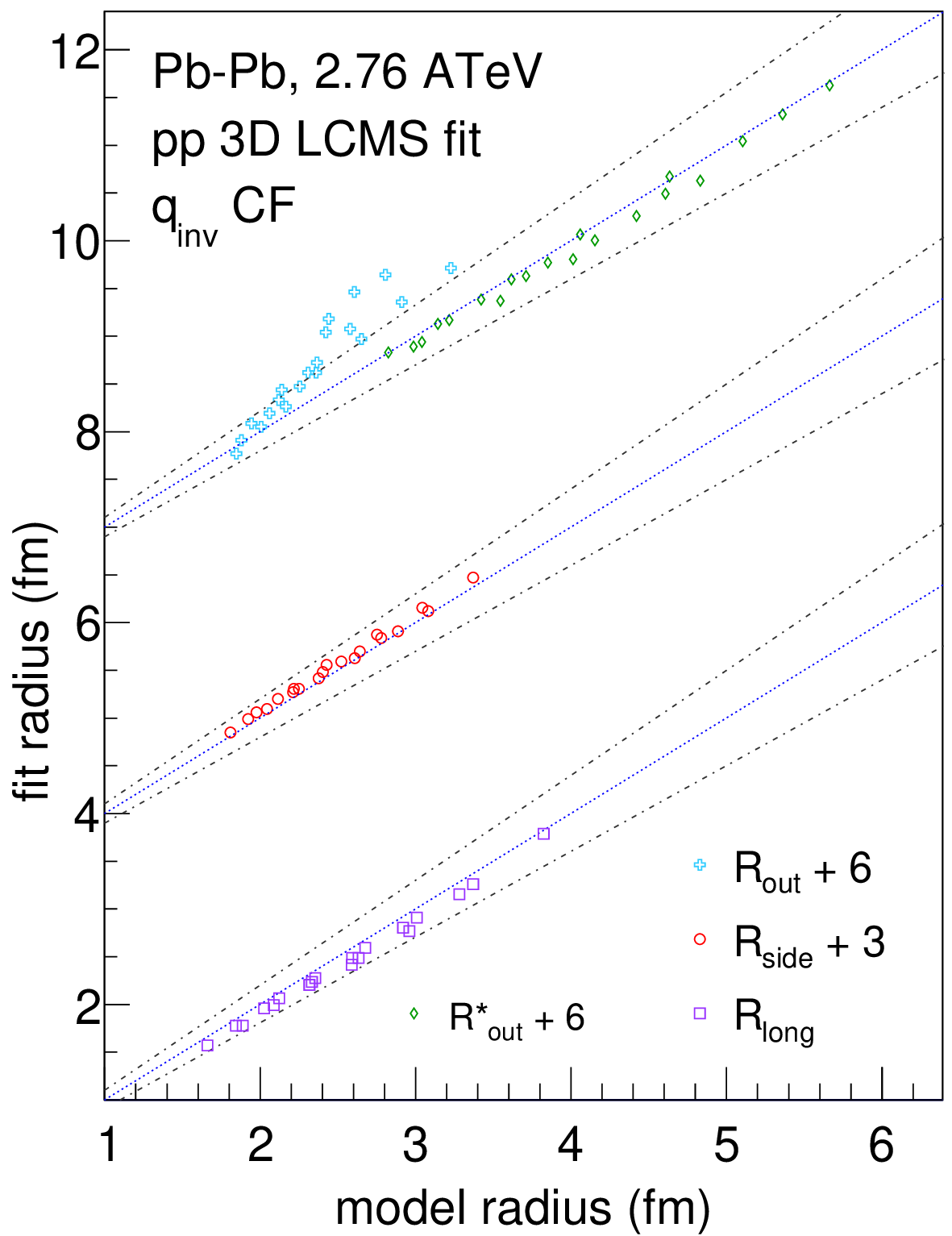}
\caption{Comparison of input model radii estimated directly from space-time separation distributions and the final fit radii in three dimensions for all centrality and pair momentum ranges. The $R^{*}_{out}$, $R_{out}$, and $R_{side}$ plots are shifted vertically for visibility. Dotted lines indicate perfect model-fit matching, while dashed-dotted line indicate a region of matching within 10\%.}
\label{fig:modelvsfitallprf}
\end{figure}

The fact that source distributions in all three directions are reproduced by the procedure in which only the correlation function is fitted is the comprehensive confirmation of the validity of the fitting procedure proposed in this work. The full validity test of the fit is shown in Fig.~\ref{fig:modelvsfitallprf}. Fit radii for all centralities and all pair momentum ranges in the "side" and "long" direction match the input values estimated directly from the model space-time distributions. The fitted "side" radius is usually higher than the input value, the discrepancy is 2.8\% on average, but is always smaller than 5\%. The "long" radius tends to be smaller than the input value, the discrepancy is 4.4\% on average, but always smaller than 7\%. In the "out" direction, the discrepancy for the radius in PRF is on average 2.8\%, but always smaller than 5\%, while the question of the radius in LCMS is discussed in the Sec.~\ref{sec:routlcmsorigin}. The remaining discrepancies might be caused by the limited numerical accuracy of the simulated correlation functions. 

\subsection{Validity of variants of the fit procedure}
\label{sec:variants}

\begin{figure}[!h]
\centering\includegraphics[width=.999\linewidth]{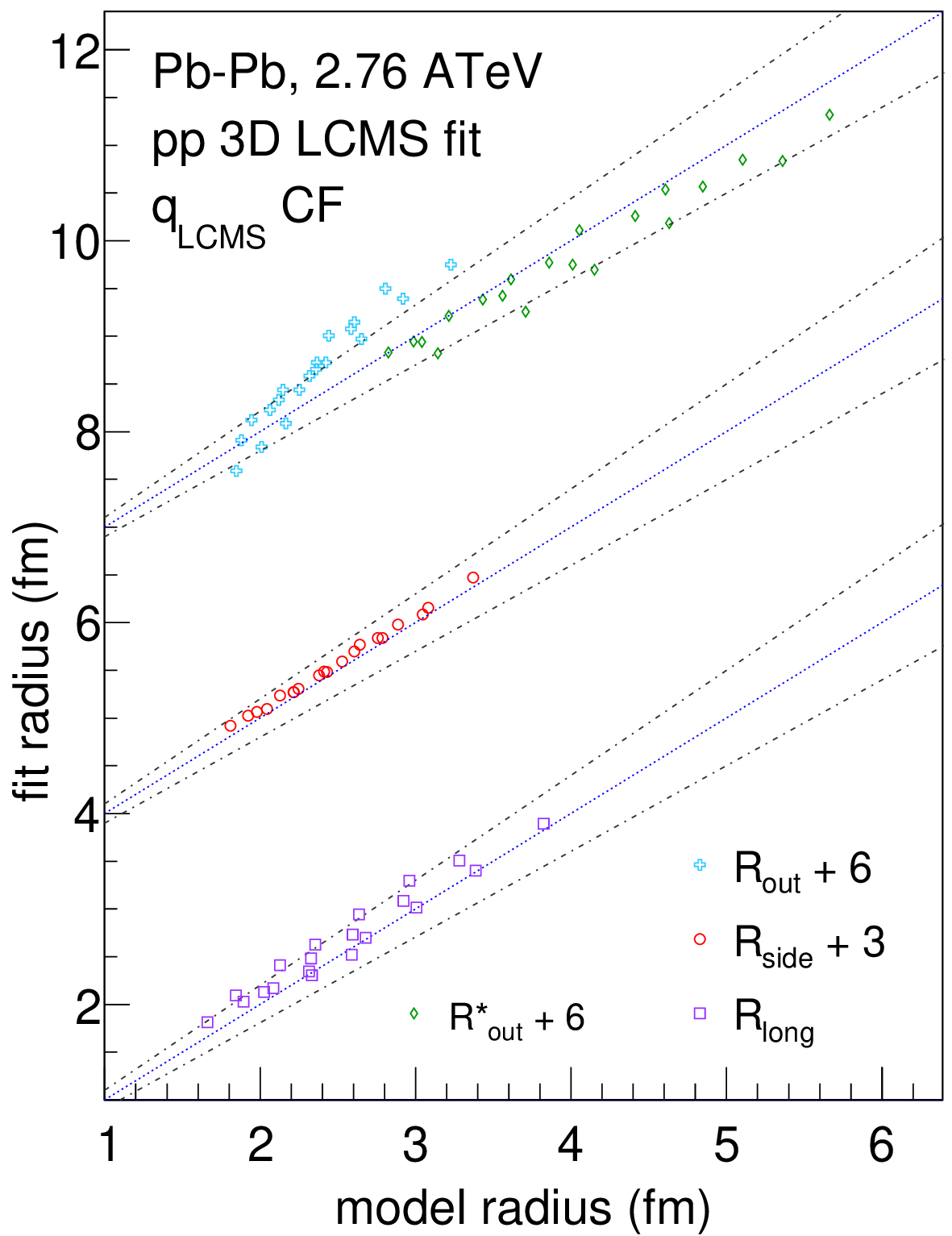}
\caption{Comparison of input model radii estimated directly from space-time separation distributions and the final fit radii in three dimensions for all centrality and pair momentum ranges. Fit is performed on correlation functions constructed vs. $q_{LCMS}$, instead of $q_{inv}$. The $R^{*}_{out}$, $R_{out}$, and $R_{side}$ plots are shifted vertically for visibility. Dotted lines indicate perfect model-fit matching, while dashed-dotted line indicate a region of matching within 10\%.}
\label{fig:modelvsfitalllcms}
\end{figure}

While the validity test of the procedure has been demonstrated in previous section, several variants of the procedure were explored to check if better matching could be achieved. As mentioned before the whole procedure has been performed independently for correlation function constructed vs. $q_{LCMS}$, instead of $q_{inv}$, in order to more closely resemble the procedure employed for pions. The validity check results are presented in Fig.~\ref{fig:modelvsfitalllcms}. This procedure also gives satisfactory results for matching the model input valued and fit results. However, the observed discrepancies are slightly larger than for the defaults procedure: the average deviation is 5.2\%, 3.1\%, and 5.5\% for "out", "side", and "long" directions respectively. Also in this modified procedure, the $R_{out}$ radius in LCMS is still containing the convolution of space and time distributions and cannot be compared directly to the model "space-only" distributions.

\begin{figure}[!h]
\centering\includegraphics[width=.999\linewidth]{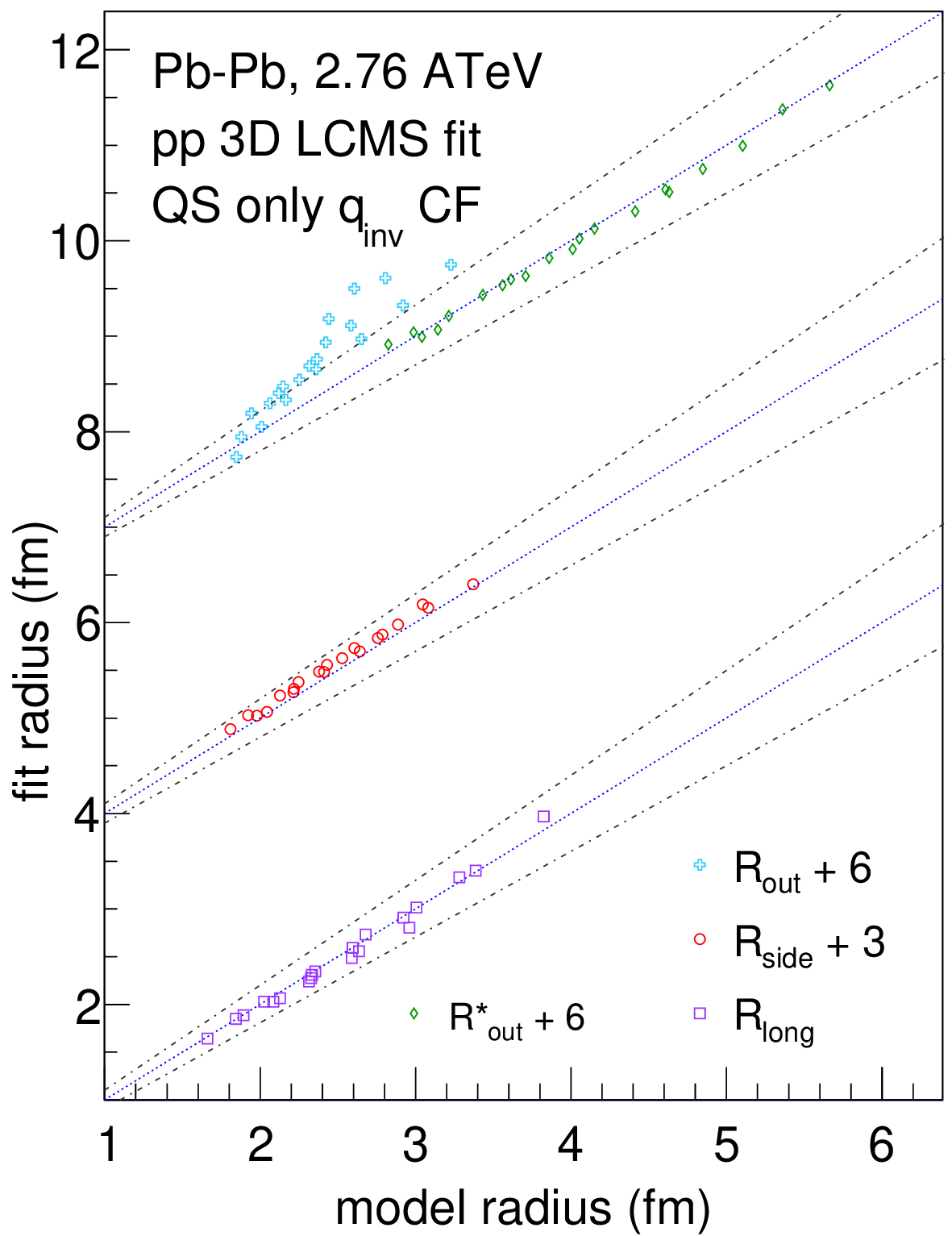}
\caption{Comparison of input model radii estimated directly from space-time separation distributions and the final fit radii in three dimensions for all centrality and pair momentum ranges. The input correlation functions and the fit calculations are performed taking into account Quantum-Statistics effects only, without the influence of FSI. The $R^{*}_{out}$, $R_{out}$, and $R_{side}$ plots are shifted vertically for visibility. Dotted lines indicate perfect model-fit matching, while dashed-dotted line indicate a region of matching within 10\%.}
\label{fig:modelvsfitallq}
\end{figure}

One might ask if the inclusion of complete FSI in the fitting procedure could be a possible source of the discrepancies between input and fitted values shown above. In order to answer this question, the full simulation and fitting procedure has been repeated for the correlation functions that include only the quantum statistics effects of (anti-)symmetrization of the pair wave function, and neglecting the FSI contribution. Such procedure cannot be reproduced in experiment, but is nevertheless a useful exercise in understanding the performance of the fit. The validity test for this procedure is shown in Fig.~\ref{fig:modelvsfitallq}. The discrepancies are slightly smaller than for the default procedure: 1.5\%, 3.4\%, and 1.8\% for "out", "side", and "long" directions respectively. Still, the effects for the $R_{out}$ radius in LCMS persist in this idealized procedure. In summary the modifications of the fitting procedure explored in this work show comparable validity as the standard procedure. Taking this into account, the default procedure constructing the correlation in $q_{inv}$ is the recommended way of performing the 3D fits of proton-proton correlations.

\subsection{Interpretation of the $R_{out}$ radius}
\label{sec:routlcmsorigin}

\begin{figure}[!h]
\centering\includegraphics[width=.999\linewidth]{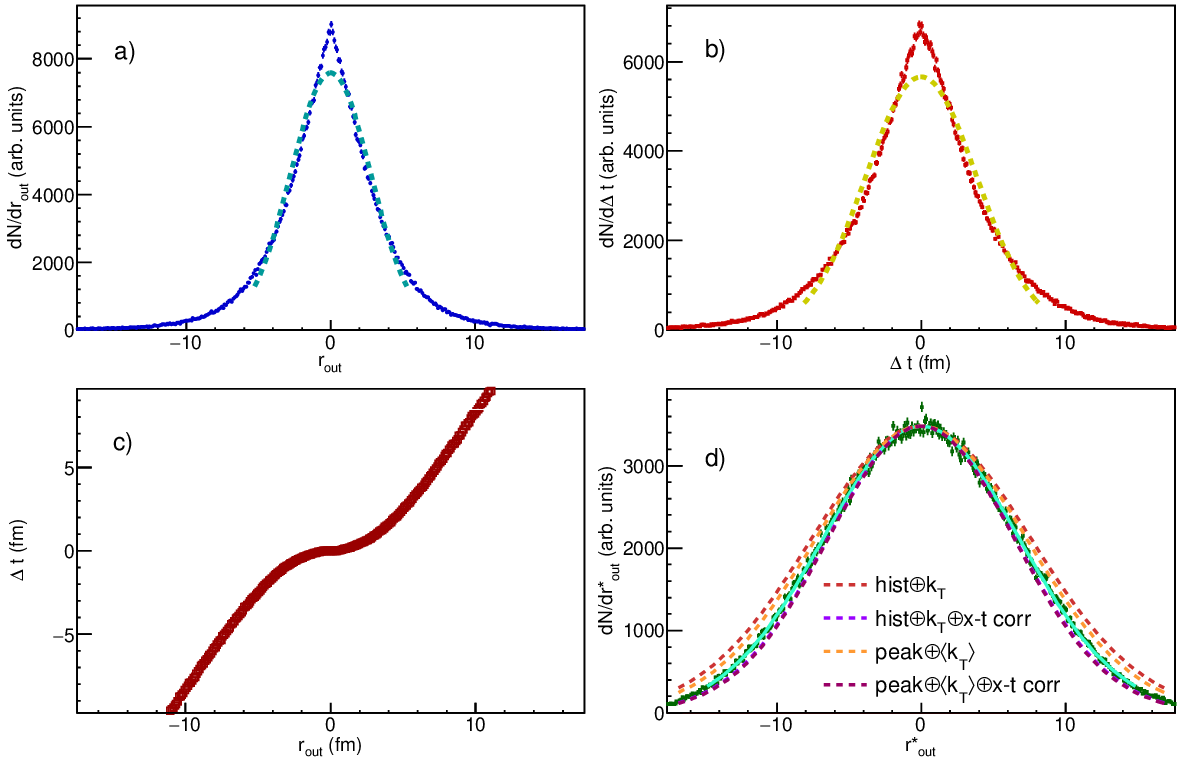}
\caption{Distributions of $r_{out}$ (a) and $\Delta t$ (b) (histograms) for an example centrality/pair momentum range, together with Gaussian fits near the peak (lines). The mean $\Delta t$ versus $r_{out}$ (panel c)). The $r^{*}_{out}$ distribution (histogram) with a Gaussian fit (solid line) (panel d)), together with selected approximations of the distribution (dashed lines) based on aggregate characteristics of the source in LCMS (see text for details).}
\label{fig:estorigin}
\end{figure}

As discussed in Sec.~\ref{sec:valid} the space-time separation of particles in PRF $r^{*}_{out}$ is a result of combing the space ($r_{out}$) and time ($\Delta t$) separation in LCMS via the Lorentz transformation with pair transverse velocity $\beta_{T}$. This formula is exact for every pair separately, but what is interesting are not values for a single pair, but rather the parameters of distributions constructed from all pairs. In particular, as shown above, the three-dimensional fit of the proton-proton correlation functions is specifically sensitive to the width of the $r^{*}_{out}$ distribution, the $R^{*}_{out}$. It is investigated if this value can be estimated based on the parameters of the distributions in LCMS - the width of the $r_{out}$ distribution - the $R_{out}$ and the width of the $\Delta t$ distribution - $R_{time}$. Top panels of Fig.~\ref{fig:estorigin} show these two distributions for one example sample for a selected centrality and pair momentum range. In addition lines on the plot show Gaussian fits to these distributions. In addition panel c) of the figure shows the space-time correlation between the two variables, which is clearly non-zero. 

Several procedures to estimate the $R^{*}_{out}$ have been investigated. Firstly, it was investigated what is the influence of the fact, that in any given pair momentum range pairs will have differing velocity, giving a range of possible values of $k_{T}$ and, therefore $\beta_{T}$. Several calculation have been performed, which have compared results where full velocity range was used with the calculations where all pairs were assumed to have the same velocity, equal to the average value in a given pair momentum range. In panel d) of Fig.~\ref{fig:estorigin} the first type of calculations are denoted with "$k_{T}$" symbol, while the one using the average is denoted with "$\langle{}k_{T}\rangle$". Secondly, the histograms shown in panel a) and b) are visibly non-Gaussian, which is true for a significant number of pair momentum ranges. It is not obvious which estimate of the width of this distribution should be used for the $R^{*}_{out}$ calculation. In one scenario, denoted as "hist", values of $r_{out}$ are generated with the probability distribution described by the full histogram directly. The second scenario is based on the fact, that QS and FSI contributions to the pair correlation function are dominated by pairs, where average separation is small. So the pairs at low $r_{out}$ may be the ones which matter most. So in this scenario, denoted as "peak", the values of $r_{out}$ and $\Delta t$ are generated according to the Gaussian, fitted in the vicinity of the peak of the respective distributions, and shown as lines in panels a) and b) of Fig.~\ref{fig:estorigin}. Thirdly, it has been investigated whether the space-time ("x-t") correlation plays an important role in the determination of $R^{*}_{out}$. Calculations in which the correlation were taken into account are denoted with "x-t corr", while calculations where space and time distributions were assumed to be independent do not have this label. 

The employed procedure was as follows. For a sample of pairs, the $r_{out}$ and $\Delta t$ values were randomly generated, according to probability distributions explained above. Then pair velocity was either generated from the allowed range, or taken as equal to the average in the range. Lastly the $\Delta t$ value was modified (or not), to account for space-momentum correlation. Then the $r^{*}_{out}$ value was calculated according to Eq.~\eqref{eq:rsovsro} and stored in the histogram. The resulting $R^{*}_{out}$ was then compared to the "true" distribution calculated directly from the model, taking into account all correlations between space, time, and pair velocity present in the simulation. Such comparison is shown in panel d) of Fig.~\ref{fig:estorigin}. The comparison shows, that variations of the procedure do produce visibly different estimates, which have various degrees of success in reproducing the "true" distribution.

\begin{figure}[!h]
\centering\includegraphics[width=.999\linewidth]{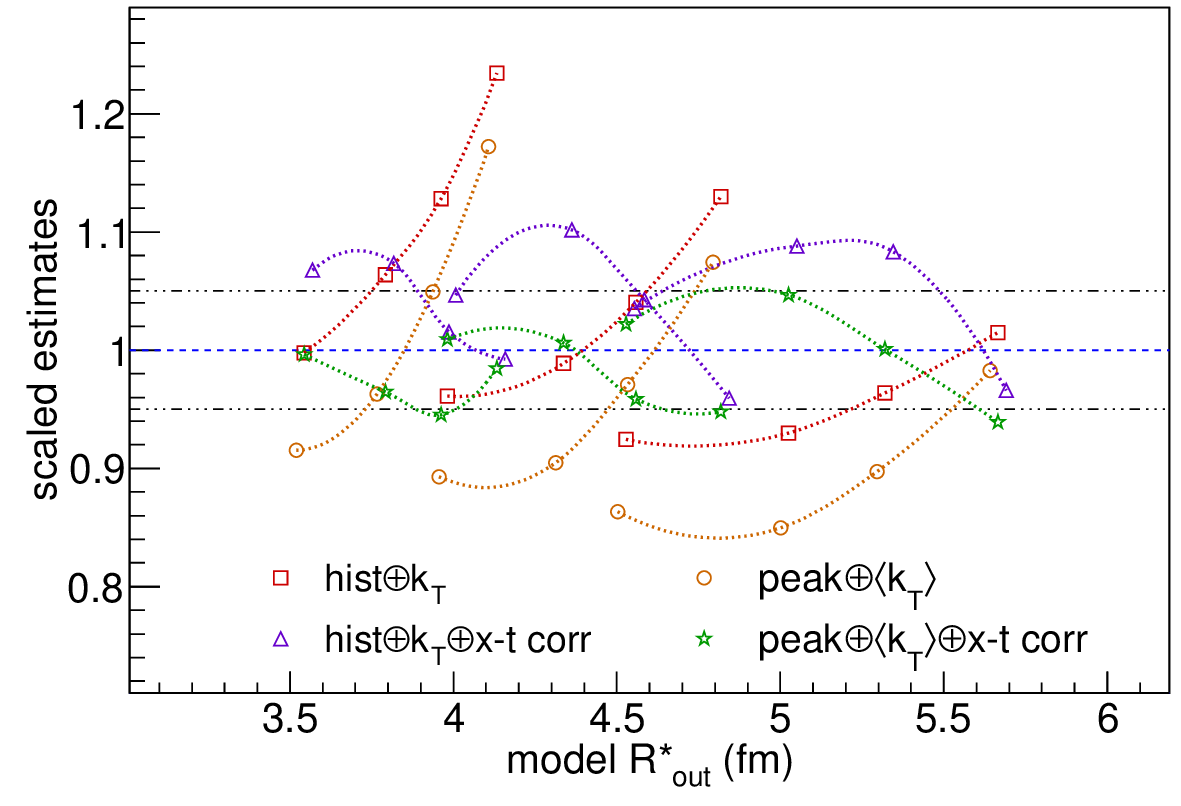}
\caption{Comparison of the width of the $r^{*}_{out}$ distribution to selected estimates based on aggregate characteristics of the source in LCMS (see text for details). The ratio between the estimate and the true value is shown for selected centralities. Lines connect estimates for four pair momentum ranges at the same centrality.}
\label{fig:estscales}
\end{figure}

The results of the comprehensive test of the estimates, for a several centrality and pair momentum ranges, are shown in Fig.~\ref{fig:estscales}. The comparison gives important insights into the factors, which are driving the final shape of the $r^{*}_{out}$ distribution. Estimates neglecting the distribution of time differences cannot reproduce the "true" values. No significant differences were found between distributions calculated using full pair momentum range and the average value. So using the simpler procedure with the average is preferred. Calculations which do not take into account space-momentum correlations are consistently worse in predicting the true value of $R^{*}_{out}$. Lastly, calculations based on the width of the distributions around the peak are consistently better than the ones which also include pairs produced further apart. 

In summary, the best estimate of $R^{*}_{out}$ is obtained in the procedure that combines: average pair velocity for the bin, $R_{out}$ and $R_{time}$ widths of the corresponding distributions around their peaks, as well as the space time correlation. With this procedure the true $R^{*}_{out}$ can be estimated with a precision better than 5\%, based on values calculated in LCMS. This conclusion is not limited only to the case of proton correlations. It is true in general, in any femtoscopic analysis performed in three dimensions, which attempts to estimate the source radii in LCMS. This includes all three-dimensional analyses to date, performed for charged pion and for kaon pairs. In particular this means, that any comparison of experimental results and model calculations for $R_{out}$ must carefully take into account all the effects mentioned above. In other words the experimental $R_{out}$ value obtained in three-dimensional femtoscopy should be understood as: "effective width of the $r_{out}$ distribution which, given an experimental average pair velocity value, produces the $r^{*}_{out}$ distribution which matches the one in data, with the assumption that emission time difference in LCMS is zero and no space-time correlations are present". It must be emphasized that correctly taking into account all of the effects mentioned above is important and not trivial. In particular it is very difficult and practically impossible to validate using single-particle model distributions alone. Therefore it is strongly recommended that any model predictions for three-dimensional radii for any type of pairs (charge pions, neutral or charged kaons, protons, etc.) is carried out by performing a full calculation of the correlation function, and extracting the radii from the fits to these calculations, following the procedure explained in this work. Several of such comparisons have been made in the past, see e.g.~\cite{Aamodt:2011mr,ALICE:2015tra,ALICE:2017gxt}, and these should be considered the most reliable data to model comparisons. While the importance of taking into account the time distribution in LCMS, and the space-momentum correlations has been recognized early in the development of the femtoscopy formalism~\cite{Chapman:1994yv}, this work quantifies the extent of this effect and gives practical prescription on how to deal with it.

\section{Summary}
\label{sec:summary}

In this work a complete formalism for a three-dimensional femtoscopic analysis for proton-proton pairs, accounting for Quantum Statistics and Final State Interaction has been developed, presented, applied to a specific model analysis and validated. The formalism consists of collecting and analyzing the three-dimensional correlation function directly in the Spherical Harmonics representation. Important components of the representation relevant for the 3D femtoscopic analysis are identified, greatly reducing the amount of data that needs to be used. A numerical Monte-Carlo procedure to generate model correlation functions is described and used to determine the best-fit parameters. The procedure exploits the flexibility of the MC approach and allows for fitting of a correlation function for any type of pair, constructed vs. relative momentum in PRF or LCMS, including any combination of QS and FSI effects. Any functional form can be used for the assumed source model, opening the possibility for fitting with non-Gaussian functions. However, for a particular task discussed in this work a Gaussian form in LCMS has been explored, which is directly comparable to 3D analyses performed to date for pions and kaons. The complete procedure is implemented in a computer code CorrFit, which can be easily run on commonly available commodity PC hardware. Tests performed in this work show that the full procedure gives precise results in realistic scenarios with reasonable requirements for computing power.

The full fitting procedure has been validated on model data. Accuracy of not worse than 5\% has been achieved in all test cases, confirming that the procedure can be considered for use in modern heavy-ion collision experiments. Special emphasis has been put on the interpretation of the "outward" femtoscopic radius. It was demonstrated that any three-dimensional analysis is only sensitive to the source distribution in PRF, while any information on the "outwards" direction in LCMS is convoluted with the time characteristics of particle emission, and can only be interpreted in this context. In particular the time and space separation in LCMS cannot be easily disentangled with three-dimensional femtoscopy. In summary this work provides a working, tested, realistic prescription for experimental analysis of three-dimensional correlation functions for proton pairs, which can now be performed at heavy-ion collision experiments.

\section*{Acknowledgments}
The author of this work would like to thank prof. Georgy Kornakov for invaluable suggestions and discussion of the results. 
This work was supported by the Polish National Science Centre under decisions  no. UMO-2022/45/B/ST2/02029.

\bibliographystyle{unsrtnat}
\bibliography{bibliography}

\end{document}